# DUST LEVITATION ABOVE THE LUNAR SURFACE: ROLE OF CHARGE FLUCTUATIONS

E.V. Rosenfeld[1], A.V. Zakharov[2]

[1] Institute of Metal Physics of Ural Branch of Russian Academy of Sciences, Kovalevskaya str. 18, Yekaterinburg 620990, Russia.

[2] Space Research Institute, Profsoyuznaya str, 84/32, Moscow, 117997, Russia.

**Abstract**

The most likely cause of levitation of dust above the surface of atmosphereless planets is the electrostatic mechanism. However, the crucial problem in the explanation of this effect is a determination of the reason why a large electric charge (units or even dozens of elementary charges) required for take-off can be accumulated on the smallest dust particles. Due to the photoeffect the charge of such value could be easily accumulated on a solitary dust particle, but if a dust particle has not yet taken off, the average value of its charge is several orders of magnitude lower because of the extremely low charge density $\sigma$ on the planet's surface. The paper shows that $\sigma$ is really small only for averaging over regions of macroscopic size, and on a submicron scale the surface appear to be a collection of chaotic "spots" with charges of different signs. The positively charged "spots" are only slightly larger than the negatively charged spots, which explains the small value of the charge density averaged over macroscopic regions. "Spots" arise due to fluctuations in the fluxes of the photoelectrons taking off and falling back the surface, and the charge density inside the "spots" is sufficient to allow a takeoff of particles with the dimensions less than or on the order of 0.1 μm in the field of a double layer.

**Key words**: dust particles, photoelectrons, charge fluctuations, levitation,

## 1. Introduction

Space bodies, devoid of atmosphere, are subject to significant influence of the solar ultraviolet radiation and the solar wind (in the absence of an intrinsic magnetic field). The upper layer of the regolith of the Moon and most small bodies can be considered as a good insulator [Carrier et al., 1991] and it is charged positively to a potential of several volts on the illuminated part of the surface, mainly as a result of photoelectron emission. Low-energy photoelectrons above the surface and the charged surface form a near-surface (about 1 *m*) electric field of ~10 $Vm^{-1}$ [Manka, 1973; Poppe and Horanyi, 2010]. It is commonly believed that in this field, submicron and micron dust regolith particles are able to lift off above the surface, forming, together with the surrounding plasma, a plasma-dust exosphere near the surface of such bodies.

The formation of the electrical potential of the surface of atmosphereless bodies is quite a complicated process, since it occurs as a result of influence of several factors - the solar wind plasma, secondary electrons, ultraviolet radiation of the sun, causing photoemission of electrons. The contribution of each of these factors to the potential of the lunar surface largely depends on the parameters of the solar wind, the angle of the Sun etc. and varies in space and time. It was shown that the photoeffect is the dominant factor in the formation of the electric potential of the illuminated lunar surface (outside the geotail) [Opik and Singer, 1960; Manka and Mitchel, 1973; Colwell et al., 2007]. The dependence of the potential of the lunar surface on the solar wind flux and photoemission under various conditions was considered in [Grobman and Blank, 1969]. An estimate of the effect of solar wind on the potential of the lunar surface was made In [Bernstain et al., 1963]. It was shown that the electrons of the quiet solar wind are the main factor affecting the potential of the surface near the terminator and on the night side of the Moon. Estimates made in [Manka and Michel, 1973] show that the potential of the illuminated surface of the moon is positive and amounts to several volts, while on the terminator it is negative and reaches several tens of volts. These estimates are consistent with measurements of the electric field near the surface of the Moon, which were performed in the SIDE experiment (Superthermal Ion Detector Experiment) [Freeman et al., 1973].

In order for an electrostatic mechanism to work the surface has to be so much charged that the electric field of the double layer arose over the surface could raise individual dust particles. However, there is still not entirely clear exactly how charges sufficient for lifting (tens or more elementary charges) on submicron dust particles in the electric field of the double layer, and how these charges are distributed between the individual dust particles on the surface. In recent years, thanks to theoretical and laboratory modeling (see, for example, [De and Criswell, 1977; Criswell and De, 1977; Poppe and Horanyi, 2010; Dove et al., 2012; Stubbs et al., 2014; Wang et al., 2016a; Vaverka et al., 2016; Popel et al, 2016] significantly advanced the understanding of the formation of the photoelectron layer. Therefore, in this paper we will focus only on one aspect of the process of formation of the potential of the illuminated surface and photoelectron layer - on the fluctuations of the charge, which in our opinion, play a fundamental role in the process the charge of the particles lying on the surface. However, we will limit ourselves to effects associated with photoemission only, neglecting the influence of solar wind plasma, secondary electrons, adhesion of particles, as well as possible the influence of the local magnetic field.

**2. Charge necessary to lift a dust particle**

The main source of the field above the illuminated surface is the double electric layer (DEL) created by photoelectrons that fly above the surface and positive "holes" remaining on the surface. Since the energy of light quanta knocking out photoelectrons should have about 10 $eV$, the potential difference $V$ within the layer has to be the same order of magnitude, since the electric field brakes the photoelectrons and returns them back to the surface (gravity in this process plays practically no role). The simplest estimates of the maximum of the lifting height of the photoelectron [Stubbs et al., 2006; Rosenfeld et al., 2016b] show that the thickness of this layer on the Moon should be of the order H~10 m, so that the field strength inside DEL $E_{DEL} \approx V/H \approx 1$ V m$^{-1}$, and the available experimental data [Freeman and Ibrahim, 1975] do not contradict this conclusion.

In order to create a field strength $E_{DEL}$ within the electric double layer (a capacitor) a charge density is required on the surface (positive electrode)

$$\langle v \rangle e = \varepsilon_0 E_{DEL} \approx 10^{-11} \text{ C m}^{-2} \quad \Rightarrow \quad \langle v \rangle \approx 6 \cdot 10^{7} \text{ m}^{-2} \tag{1}$$

where $\langle v \rangle$ is the number of excess charges +e per 1 m².

Since the field $E_{DEL}$ controls the return of all knocked out photoelectrons from the surface, the time of DEL formation is the time during which $v$ electrons are knocked out from each square meter of this surface. Therefore, if the density of the photocurrent in the lunar midday zone [Feuerbacher et al.,1972]

$$j_{ph} \approx 5 \text{ μA m}^{-2} \approx 3 \cdot 10^{13} e \text{ m}^{-2} \text{ s}^{-1} \tag{2}$$

time of the order $ve/j_{ph} \approx 2$ μs requires for the formation of a double layer. At the end of this time, $v$ positively charged holes will be more or less homogeneously distributed for each square meter of the surface. Therefore, the average charge of a dust particle with a diameter $d$ lying on the surface (or any other area of the surface with the same diameter), i.e. the share of excess charges +$e$ attributable to it becomes and remains equal to

$$\langle q(d) \rangle = \frac{1}{4} \pi d^2 v e \approx 5 \cdot 10^{7} d^2 e \tag{3}$$

From (3) follows that $\langle q(d) \rangle$ can exceed one (!) elementary charge of *e* only if the diameter of a dust particle not less than 140 *microns*. For smaller dust particles, the probability (proportional to $d^2$) to acquire this minimally possible charge is much less likely: in the mean only one of the 2 $10^6$ of dust particles with a diameter *d = 100 nm* has a charge *e*, and one of the 2 $10^8$- with a diameter *d = 10 nm*.

Now let's calculate which charge for a dust particle with a diameter of *d nm* is necessary, that a field with a strength of E can raise it on the Moon. This becomes possible as soon as the Coulomb force exceeds the force of gravity, but it must also be taken into account that the charge must be equal to the integer number of *e*. Consequently, the minimum number of electrons $n_{min}$, which a dust particle of diameter *d* has to lose for take-off is equal:

$$n_{min}(d) \approx \left\lfloor \frac{\rho g d^3}{2eE} \right\rfloor + 1 \approx \left\lfloor 1.5 \cdot 10^{22} \frac{d^3}{E} \right\rfloor + 1 \qquad (4)$$

where $\lfloor x \rfloor$ means the integer part of *x*, $g \approx 1.6 \, \text{ms}^{-2}$ is the acceleration of free fall on the Moon, and $\rho \approx 3000 \, \text{kg m}^{-3}$ is the density of the lunar regolith [Vaniman et al., 1991]. Thus, with *E~1 Vm$^{-1}$* the one electron charge is sufficient to raise a dust particle with a diameter no more than 40 *nm*, while for a particle 100 *nm* requires a charge $15e$, and for micron-sized particles - $15000e$ etc.

### 3. Problem of mean values

Thus, if we assume that all excess charge available on the surface is the charge of holes, the number of which is equal to the number of photoelectrons soaring above the surface, we come to the sad conclusion that the electrostatic mechanism is able to control the rise above the surface only a vanishingly small fraction (not more than 1/10,000,000) of particles of the fine fraction (the size less than 30 - 40 nm). If so, we would have to conclude that the electrostatic mechanism is completely ineffective, and levitation of dust above the surface is not connected with the electric field. Therefore it is necessary either to find the reason why dust particles at very low average surface charge density, still could obtain a charge sufficient for takeoff, or to look for other, not the electrostatic mechanism of a levitation of dust.

As for an alternative to electrostatic mechanisms, the only example known to us considered in [Rosenfeld et al., 2016a] is the rise of dust particles from the surface due to thermal fluctuations. In this paper, it is shown that the "boiling layer" of dust with a thickness of a few decimeters can form above the hot (about 400K [Vaniman et al., 1991]) surface of the Moon. This layer consists of dust grains with a diameter of the order of 10 *nm* which have the thermal velocity sufficient for levitation at an altitude of about a meter for several seconds. In contrast to the electrostatic mechanism capable to lift only charged dust particles, thermal fluctuations lift up almost all fine particles lying on the illuminated surface. However, this mechanism cannot explain neither the levitation of a larger (50 -100*nm* and more) dust, nor especially higher-altitude levitation [Stubbs et al., 2006, 2014].

Back on electrostatic mechanisms, we note that a charge sufficient to lift a dust particle could easily accumulate due to the photoelectric effect on a solitary particle [Walch at al., 1994]. In our case the dust particles recumbent on the surface have to get charges and so it is necessary to use formulas of the previous section, but not formulas for a field of point charge (see also the "isolated capacitors" and the "shared charge" models in [Flanagan and Goree, 2006)]). In addition, it is clear that any attempt to find a mechanism that is able to increase the average charge *q* of a dust particle cannot give the desired result. Indeed, as *q* increases by a factor of N, the field above the surface should be increased on the same value, or the already vanishingly small number of charged dust particles would decrease N times at the same value of $E_{DEL}$.

A more sophisticated possibility of a dramatic increase in the charges of dust particles was considered by the authors of [Wang et al., 2016b]. They pointed out that light quanta or fast particles can penetrate into the gaps between the dust particles of the uppermost layer of the surface and knock out electrons from the underlying particles. The authors of [Wang et al., 2016b] indicated that these photoelectrons will be absorbed by dust particles of the upper layer, and the growth of positive and negative charges should continue until the potential difference between them reaches a value *V* ~ 10 V that will cause the emitted photoelectrons to return back. With submicron sizes of dust particles (and therefore similar sizes of cavities between them), this requires fields with very high intensity and, consequently, high charge densities on the surfaces of dust particles. It seems to us, however, that this mechanism cannot give the desired effect. Indeed, although the dust particles of the upper layer can receive a significant negative charge, but the dust particles of the lower layer - a positive charge of the same in magnitude. In this closed system, almost all the electric field will be concentrated in the cavities between these two layers of dust. Moreover, unlike charged dust particles are strongly attracted to each other, creating more or less large neutral clusters with a mass exceeding the mass of a single dust particle, so that the external field will not be able to lift them.

Despite this, we believe it is extremely important to pay attention to the assumption underlying the model of [Wang et al., 2016b; Zimmerman et al., 2016]. An important and essentially new step in this model is the fact that free charges of both signs can appear in the surface layers of dust, and the densities of these charges can significantly exceed the mean density (1). It is critically important, however, that at the same time the nearest neighbors of the charged dust particle have charges of the same sign, and the oppositely charged dust particles lie on the surface farther from it. In this case, Coulomb repulsive forces would exceed the forces of attraction, and there would be a real possibility of forming an electrostatic mechanism of dust levitation. If the indicated charge densities were indeed large enough, this could explain the mechanism of the appearance of large charges of individual dust particles on the surface, which are necessary for levitation.

Therefore, in this paper, we will consider the fluctuations of charge density arising on a surface on which a light quantum flux creating a positively charged holes and the flux of photoelectrons returning back fall down at the same time. If the light continues not too long, it is extremely unlikely that the falling electron enters the atom with the "hole" that arose previously due to the photoelectric effect. Consequently, the total value of both the positive and negative charges accumulating on the surface is proportional to the duration of the illumination, and these opposite charges are randomly distributed over the surface without annihilating each other (the conductivity of the regolith in the absence of water even on the illuminated surface of the moon is small [Carrier III et al., 1991]).We also try to simulate in such process the dynamics of the development of fluctuations of the surface charge density in this work. We emphasize that in this case, unlike the model [Wang et al., 2016b; Zimmerman et al., 2016], opposite charged regions appear to be distributed over the surface, so that the forces of attraction between the dust particles are mainly horizontal, and the repulsive forces are mainly vertically directed, which ensures the appearance of a lifting force.

**4. "Random walks" of the charge of dust particles**

As mentioned above, for some time after the occurrence of the double layer the photoelectrons falling back to the surface almost certainly fall into neutral atoms and not into those atoms from which the electron was previously knocked out. This means that the positive and negative charges $\pm e$ are distributed randomly over the surface, but with an almost identical and proportional to light time $t$ average density

$$\nu(t) = j_{ph}t/e \approx 3 \cdot 10^{13} t \text{ m}^{-2} \qquad (5)$$

The difference between the mean values of the number of positive and negative charges in any area of diameter $d$

$$\langle n_+(d,t)\rangle - \langle n_-(d,t)\rangle = \frac{\langle q(d)\rangle}{e} \approx 5\cdot 10^7 d^2 \qquad (6)$$

is absolutely insignificant against the background of the linearly increasing number of charges of each sign in this area

$$\langle n_+(d,t)\rangle \approx \langle n_-(d,t)\rangle \approx \langle n(d,t)\rangle = \pi d^2/4\, v(t) \approx 2\cdot 10^{13} d^2 t \qquad (7)$$

($t$ in *seconds*, $d$ in *m*).

Although the average values **of** $n_+(d,t)$ and $n_-(d,t)$ grow with time linearly, their true values change randomly, since the drop of both electrons and light quanta, like the emission of a photoelectron, are purely random processes. Therefore, there must be "random walks" of the charge (see e.g. [Weiss and Rubin, 1982]) - the charge of any area must constantly change not only the magnitude but also the sign. One "step of walk", i.e. change of a charge on $\pm e$, on average takes time

$$\delta t(d) = 4e/\pi d^2 j_{ph} \approx 4\cdot\left(10^{-7}/d\right)^2 \text{ s} \qquad (8)$$

This is of the order of 0.04 *s* for the micron area, 4 *s* for the area with a diameter of 100 *nm* and about 6 - 7 minutes if its diameter is 10 *nm*. As a result of such "walks" a few minutes after the sunrise and the appearance of the photocurrent, the charge of a particular sign should appear not at a millionth of a percent, but almost on every dust particle with a diameter of 10 *nm* or more. However, the total amount of dust particles with an excess of electrons is only slightly less than with their deficiency, so the sum of charges of dust grains with different signs almost exactly compensate each other. As a result of it the average value of the charge density (1) on the surface is so small in comparison with the charge density of each sign (5). For the same reason, the average value of the charge (6) of any area (or dust particle) is so small in comparison with the charge of each sign (7) accumulated in this area during a very large (in comparison with $\delta t(d)$ (8)) time interval.

Further, since the values $n_+(d,t)$ and $n_-(d,t)$ change randomly, more and more noticeable fluctuations in the magnitude and the sign of their difference $\delta n(d,t) = n_+(d,t) - n_-(d,t)$ should appear with time. At the same time, the average value $\langle \delta n(d,t)\rangle$ (6) remains negligibly small (see Section 2). It is well known (see, for example, [VanKampen, 2007] and references therein) that for purely random walks the root-mean-square deviation increases in proportion to the square root of the "number of steps". In our case, for an area (or for a single dust particle) with a diameter $d$, the number of steps taken in a time $t$ is a ratio $t/\delta t(d)$ or simply $n(d,t)$ (7). Therefore, the quantity $\sigma(d,t)$ determining the value of the number of excess elementary charges averaged over such regions (dust particles) at time *t* is determined by the formula

$$\sigma(d,t) = \sqrt{\langle [n_+(d,t) - n_-(d,t)]^2 \rangle} \approx \sqrt{n(d,t)} \approx 5 \cdot 10^6 d\sqrt{t} \qquad (9)$$

(time in seconds). Comparing $\sigma$ and $n_{min}(d)$ (4), we get a relation between the diameter of a dust particle and the time it takes to accumulate a charge sufficient for take-off:

$$n_{min}(d) = \sigma(d,t) \quad \Rightarrow \quad t_{rise} \approx 10^{-5} d^4 \qquad (10)$$

(time in seconds, diameter in nanometers). It follows, in particular, that for such purely random "walks" of a charge it will take about a minute between the start of the illumination of a particle with $d \approx 40$ *nm* and the instant when it emits a photoelectron and take-off; for a particle with $d = 100$ *nm* it will take about 20 minutes to accumulate the charge necessary for take-off; and for $d = 200$ *nm* - about five hours. Note that for the smallest dust grains with $d < 40$ *nm* the formula (10) does not work: such particles must emit a single photoelectron for take-off, and the probability of this is proportional to $d^2$, see (8). This means that almost immediately after sunrise, dust particles with a diameter of about 40 *nm* fly off, and all the rest – the later, the more their diameter differs from 40 *nm*.

Since, according to formula (9), the amplitude of the charge fluctuations increases monotonically with time, it may seem that a charge sufficient for take-off should accumulate with time on almost half of all the dusts on the surface (in the second half the charge will be negative). In reality, however, there are two factors that prevent it. First, when the density of elementary charges on the surface $\nu(t)$ (5) approaches the density of the atoms forming it (for near 0.5-*nm* interatomic distances it is about $5 \cdot 10^{18}$ m$^{-2}$), a significant part of the electrons falling back to the surface will annihilate with the holes, and $\langle n_\pm(d,t) \rangle$ (7) will cease to grow. Thus, the atomic structure of matter sets a natural limit on the accumulation of different surface charges. Secondly, there must always be a mechanism for suppressing fluctuations, because of the action of which their lifetime, and therefore their amplitude, are limited. In our case, this is the Coulomb mechanism discussed in the next section. This mechanism relates to the fact that the positively charged areas attract, and the negatively charged ones - repel the photoelectrons falling back on the surface.

## 5. The attenuation of the charge fluctuations

The slow but unlimited increase with time of the average value of the modulus of the accumulated charge will actually occur only if the probabilities of knocking out of an electron of the area and it falling into the same area are the same. This is true for a neutral area (or a dust particle), but when uncompensated charges appear, the situation changes. *Ceteris paribus* the probability of an electron falling into a positively charged area must be higher than the probability of photoemission(charge independent), and vice versa for a negatively charged one. This means that the "walks" of the charge ceases to be random, because a constant force appears, tending to return the total charge of the area to zero. Of course, computer simulation could give the most accurate view of the dynamics of surface charge fluctuations, but we will try to consider some principal lows of this process in an analytical form, using the methods of the theory of the stochastic differential equations (Øksendal, 2000).

For this purpose we will imagine a plane on which randomly distributed approximately the same quantities of positive and negative charges. In general, we get something like a chessboard with crooked cells (areas) that have a positive or negative charge according to their color. Near the plane - at an altitude $h$ that is less than or comparable to the characteristic cell size $d$ - the electric field created by this charge distribution will be extremely inhomogeneous. However, at a greater distance from the surface $h>>d$, a plurality of areas with charges of different signs produce approximately equal in magnitude, but opposite in direction fields. As a result, as $h$ increases, the random contributions are more and more exactly compensated and for $h >> d$ the field becomes a field of a uniformly charged plane $E_{plane} = Q/(2\varepsilon_0 S)$ ($S$ is the area of the plane, $Q$ is its total charge).

If we further simplify the situation and consider the ideal case, when all the cells (areas) on the surface are squares of the same size, having the same charges on the modulo, it can be shown that the intensity of the field created by them will decrease exponentially with distance from the plane [Rosenfeld, 2000]. This means that the component of the field strength $E_{\parallel}$ parallel to the surface (it is maximal near the surface close to the boundaries between the oppositely charged areas) practically disappears with altitude where a height **is** several times larger than the dimensions of the fluctuating areas. The component of the field perpendicular to the plane $E_{\perp}$ is directed upward above the positively charged regions, downwards over the negatively charged ones, and is maximum in magnitude near their centers. With increasing $h$, the inhomogeneities of this component also rapidly disappear, and approach to a uniform value $E_{plane}$.

The most important role of the component $E_\perp$ is that it creates the force that detaches the flying particles from the surface, but in addition, it slows down or accelerates the electrons falling on the surface in the last segment of the flight. However, in the simplest approximation, this change of the velocity can be neglected in comparison with the speed $v = \sqrt{2eV/m_e}$ that the photoelectron returning to the surface picks up inside DEL.

The $E_\parallel$ component of the field plays a much more important role in the curvature of electron trajectories near the surface. It is this component of the field that deflects the trajectories of the falling electrons from the negatively charged sections and drawing them to positively charged areas (particles).

To clarify this, Fig. 1 presents an idealized picture of the trajectory sections of photoelectrons when they fall onto the surface. At high altitude $h >> d$ vertically falling electrons move along straight lines, which are called "sighting lines" in mechanics. The electrons flying towards the negatively charged sites will get to them only if their sight lines pass through the ones indicated in Fig. 1 as the BN ("bottlenecks") area, with a cross-section $S_{BN}$. The trajectories of electrons passing through these BNs deviate from the centers of the negatively charged regions not too much, and therefore will not go beyond them. Otherwise, the field $E_\parallel$ will divert their trajectories so strongly that they will fall already on the positively charged areas.

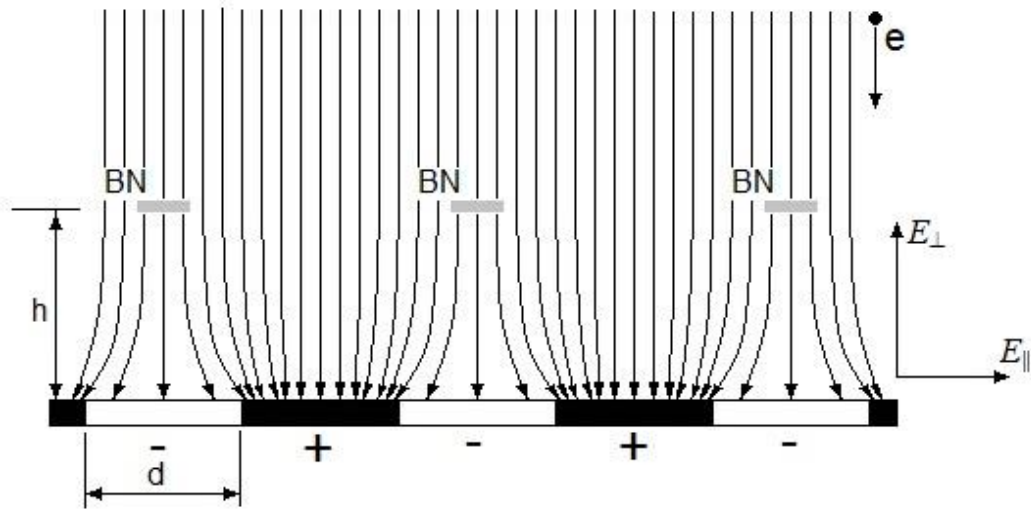

Figure 1. The uniform flow of electrons vertically falls on the plane, where there are oppositely charged (dark and light) areas. Only those electrons whose sight lines pass through the BN areas located above the centers of the negatively charged areas on the surface can enter the surface of the negatively charged regions. The remaining electrons are deflected by the $E_\parallel$ field component and fall onto the positively charged areas.

Thus, the probability of an electron to fall into an area (each such hit changes its charge by -e) depends on the charge of this region. At the same time, the probability that a photon falls into this region and an additional charge + e appears in it does not depend on the magnitude and sign of the surface charge. Consequently, at the average the same number of holes $Sj_{ph}/e$ is produced on any part $S$ of the surface in a second, but the number of electrons falling on this area depends on its charge. It is clear that the direction of the velocities of the lifting-off photoelectrons are randomly distributed around the direction of the normal to the surface. Therefore, the electrons emitted from any (but not too far removed) neighboring area can fall on any site. This means that on average, the same number of electrons directed to (but not necessary fall into) any area per a second as many as it takes off, but the fluctuations in these flows arise independently of each other.

For simplicity, we assume that the uniform electron flux falls strictly vertically at a height $h >> d$, and the electron trajectories are curved only near the surface where the field $E_\parallel$ works on, see Fig. 1. Then only the $S_{BN} j_{ph}/e$ electrons will fall on the negatively charged area $S$ by a second, the remaining $(S - S_{BN}) j_{ph}/e$ electrons directed to it will be deflected by the $E_\parallel$ field and will fall on neighboring positively charged sites. If, as we have assumed above, the areas of positively and negatively charged sites are the same, then as many extra electrons arrive in each positively charged region as they do not reach the negatively charged area. Taking this into account, the rate of change in the number of elementary charges in any area is

$$\frac{d\delta n}{dt} = -j_{ph}\Delta S/e \; \text{sign}(\delta n), \quad \Delta S = S - S_{bn}. \tag{11}$$

In order to obtain at least a rough estimate of the magnitude $\Delta S$, let us imagine that the entire charge $q = \pm e\delta n$ of each square "cell" with edge $2r$ is concentrated at its center. Then on the middle of the edge, which is the boundary of two "cells" with charges of different signs

$$E_\parallel \approx 2\frac{e}{4\pi\varepsilon_0 r^2}\delta n \tag{12}$$

(the coefficient 2 here roughly takes into account the fact that two neighboring opposite charged areas create equally directed fields $E_\parallel$ on the boundary).

Suppose that the intensity of this component of the field first changes slowly with height, and then decrease sharply to zero at an altitude $h$ (equal to a few $r$). Then the time during which the field (12) accelerates the electron is of the order $h/v \approx h\sqrt{m_e/2eV}$, and the distance by which it moves

$$\Delta r \approx \frac{at^2}{2} \approx \frac{h^2}{8\pi\varepsilon_0 V r^2}\delta n. \quad (13)$$

It is clear that at different points of an edge separating cells with charges of different signs, the field will be different, in particular, in the corners of the cells it disappears. In order not to take this into account, we extremely simplify the situation and assume that our areas have the form of a circle with radius $r$, and $E_\parallel$ on its boundary is defined by formula (12). In this case, $\Delta S$ is easy to find:

$$\Delta S \approx 2\pi r \Delta r \approx \frac{h^2}{4\varepsilon_0 V r}\delta n \quad (14)$$

Of course, this is an extremely rude approximation, although it can be assumed that the proper choice of the value of the adjustable parameter $h$ should minimize its limitations. We emphasize once again that we are not trying to obtain exact quantitative results here, and we are mainly interested in general qualitative conclusions. The approximation (14) is introduced simply because it allows us to obtain finite formulas in an analytical form. Substituting (14) into (11), (13), we arrive at the standard equation of the type of the radioactive decay equation:

$$\frac{d}{dt}\delta n = -\frac{h^2 j_{ph}}{2\varepsilon_0 V d}\delta n, \quad (15)$$

This equation predicts that the magnitude of the excess charge will decrease exponentially with time:

$$\delta n = \delta n_0 \exp(-t/\tau), \quad \tau = \frac{2\varepsilon_0 V d}{h^2 j_{ph}} = \frac{\varepsilon_0 m_e d}{h^2 j_{ph}}v_0^2, \quad (16)$$

where $v_0$ is the speed of the rise / fall of the photoelectron.

It is important to note here that the relaxation time $\tau$ is proportional to the square of the electron velocity. The reason is that the higher their velocity, the shorter the time during which the field $E_\parallel$ deflects them, and therefore the smaller the distance that they will be deflected. This leads to an important conclusion: the higher the velocity of the incident electrons, the slower the charge fluctuates will be "dissolved", and therefore this fluctuation can reach a higher maximum charge.

In fact, however, besides this relatively slow unidirectional change of charge with time, fluctuations will arise, because both the drop of electrons and light quanta on the surface, and the emission of photoelectrons occur purely randomly. Therefore, in the next section, an attempt is made to describe the charge density fluctuations on a uniformly illuminated surface using the theory of stochastic differential equations [Øksendal 2000].

**6. Parameters describing the evolution of the charge fluctuations**

Stochastic differential equations are used in describing systems in which the increment $dx$ of some physical quantity $x$ in time $dt$ is determined not only by the usual smooth processes occurring in the system, but also by fluctuations:

$$dx = a(x,t)dt + fluct(x,t,dt). \qquad (17)$$

The first term here describes the usual smooth change occurring at a rate $a(x,t)$ and proportional to $dt$. Considering the Wiener process, i.e. a continuous random walk (the duration and length of the step tend to zero), we have to write

$$fluct(x,t,dt) = \sigma(x,t)\delta W, \quad \delta W = \varepsilon\sqrt{dt} \qquad (18)$$

where σ is standard deviation (SD), and $\varepsilon$ is a Gaussian random variable with a zero mean and a unit variance.

The solution of the equations (17), (18) is not one curve, but the array of an infinite number of such curves. Each of them describes the behavior of one of the (identical) systems of the ensemble originally located in the same initial states and evolved under the same conditions. In this case, measuring the value of $x$ for each system at time $t$, we get the Gaussian distribution with the mean $\langle x(t) \rangle$ and variance $\sigma^2(t)$. The value of $x$ for each individual system will then differ from $\langle x(t) \rangle$ by a random amount $\sigma(t)\cdot\varepsilon$, so that the graphs of the values of $x(t)$ for different systems (in our case - areas) will lie mainly near the line $\langle x(t) \rangle$ - inside a neighborhood with a width of order $2\sigma(t)$.

The theory of stochastic differential equations [Øksendal 2000] allows us to calculate the time dependences of the mean value and SD for the sought value (in our case this is $\langle \delta n(t) \rangle$ and $\sigma(t)$) if there are known the time dependences of the coefficients $a(\delta n,t)$ and $b(\delta n,t)$ before $dt$ and before the infinitesimal increment of the Wiener noise $\varepsilon\sqrt{dt}$ in the equation (19)

$$d(\delta n) = a(\delta n,t)dt + b(\delta n,t)\varepsilon\sqrt{dt} \qquad (19)$$

This equation is not an ordinary differential equation, it has a formal nature, and $a(\delta n,t)$ and $b(\delta n,t)$ mean here the rates of change $\langle \delta n \rangle$ and SD at the time *t*. In our case, these quantities have already been determined (see (15) and (9)):

$$a(n,t) = -\frac{h^2 j_{ph}}{2\varepsilon_0 V d}\delta n \approx 3\cdot 10^4 \, p^2 \delta n \cdot d \ \text{s}^{-1}, \quad b(n,t) = \sqrt{2\frac{\pi d^2 j_{ph}}{4e}} \approx 7\cdot 10^6 d \ \text{s}^{-\frac{1}{2}} \tag{20}$$

The factor 2 under the root in the formula for *b* takes into account that the flux of emitted electrons and the flux of incident electrons fluctuates independently. In addition, a notation $p = h/d$ is introduced here. This value shows how many times the height *h* at which the component $E_\parallel$ of the electric field disappears is larger than the size of the fluctuating region *d*. Since *h* increases with increasing *d*, we can assume that the dependence $p(d)$ is weaker than the dependence $h(d)$. The parameter *p* is several units in the order of magnitude, but it is hardly possible to specify a more precise value here.

Substituting these results in (19), it is natural to neglect the extremely small average value of the number of elementary charges $\langle \delta n \rangle = \langle q(d) \rangle / e$ (see (3) and the text relating to this formula) in comparison with its instantaneous value $\delta n(t)$. As a result, we obtain a standard equation describing the Ornstein-Uhlenbeck process

$$d(\delta n) = -\delta n \frac{dt}{\tau} + \sigma_0 \delta W;$$
$$\tau(d) = \frac{2\varepsilon_0 V}{p^2 d j_{ph}} \approx \frac{3\cdot 10^{-5}}{p^2 d} \ \text{s}, \quad \sigma_0(d) = d\sqrt{\frac{\pi j_{ph}}{2e}} \approx 7\cdot 10^6 d \ \text{s}^{-\frac{1}{2}}. \tag{21}$$

(*p* is dimensionless, *d* in meters). The solution of this equation [Øksendal 2000] has the form

$$\delta n(t) = \delta n_0 \exp(-t/\tau) + \sigma_\infty \sqrt{1 - \exp(-2t/\tau)}\, \varepsilon,$$
$$\sigma_\infty(d) = \sigma_0(d)\sqrt{\frac{\tau(d)}{2}} = \frac{d^{\frac{1}{2}}}{p}\sqrt{\frac{\pi\varepsilon_0 V}{2e}} \approx 3\cdot 10^4 \frac{d^{\frac{1}{2}}}{p}. \tag{22}$$

Accordingly, the spectral function of the fluctuations (that is, the Fourier transform of the autocorrelation function) is a Lorentz function

$$S(\omega) \equiv \frac{1}{\pi}\int_{-\infty}^{\infty} \text{cov}(t)\exp(i\omega t)\,dt = \frac{1}{\pi}\frac{\sigma_0^2}{\omega^2 + \tau^{-2}}. \tag{23}$$

Here we use the standard notation

$$\text{cov}(t_1, t_2) = \langle (\delta n(t_1) - \langle \delta n(t_1) \rangle)(\delta n(t_2) - \langle \delta n(t_2) \rangle) \rangle$$

and it is taken into account that $\text{cov}(t_1, t_2) = \text{cov}(t_1 - t_2)$ in the case of a stationary process.

In our case, the meaning of the spectral function $S(\omega)$ is very simple. Since the function $\delta n(t)$ (22) describes the time dependence of the excess charges in the area under consideration, its sign changes varies with time and $\delta n(t)$ can be represented as a sum of sinusoids with different periods $T = 2\pi/\omega$, and $S(\omega)$ determines the probability that a fluctuation with frequency $\omega$ will be represented in this expansion. Thus, $S(\omega)$ determines the characteristic range of frequencies of change oscillations in the area. It can be seen from (23) that all low-frequency oscillations (with a period exceeding $2\pi^{3/2}\tau(d)$) are approximately equally probable, but the probability of more rapid oscillations falls off rather sharply with the frequency.

Further, as seen from (22), the characteristic lifetime of the fluctuation $\tau$ determines the time scale of two processes at once. On the one hand, this is the time during which the number of superfluous / missing electrons becomes $e$ times less than its initial value $\delta n_0$, and on the other hand - just this order of time must pass after the start of illumination, so that the amplitude of the charge fluctuations (i.e., SD=$\sigma$) has acquired a constant value $\sigma_\infty$. It follows from (21) that $\tau(d)$ is inversely proportional to the size of the fluctuating area $d=2r$ (or the diameter of the dust particle). The reason is very simple: the area $\Delta S$ (14), which determines the rate of disappearance of the charge of fluctuation, is proportional **just** to the linear size $d$ of the area of fluctuation. As a result, the lifetime $\tau$ (21) of fluctuations with characteristic dimension $d \sim 10$ *nm* is several minutes, if $d \sim 100$ *nm* $\tau$ is of the order of tens of seconds, and for micron areas – fractions of a second.

It is also easy to understand the meaning of the expression (22) for the steady-state value of the amplitude of the charge fluctuations $\sigma_\infty(d)$ (it is the asymptotic value of $\sigma$ at $t >> \tau$). This value, which determines the **variation** of the number of excess charges $\delta n = n_+ - n_-$ from one area to another, was previously discussed in Section 4. According to the formula (9) the standard deviation $\sigma$ is proportional to $\sqrt{t}$ for purely "random walks", when the increase and decrease of the charge under fluctuations are purely random. If the regular processes (15) limit the lifetime of the fluctuations, then the charge increases only during a time of the order of $\tau(d)$, and we must obtain from (5), (7) and (9)

$$\sigma(d) \leq \sigma_{max}(d) \sim \sqrt{\pi j_{ph} d^2 \tau(d)/4e}. \qquad (24)$$

Taking into account the expression (21) for $\tau(d)$, we see that $\sigma_{max}(d) \sim \sqrt{d}$, and the formula (22) gives the same result but with the correct coefficient.

Comparing $\sigma_\infty$ (22) with the number of elementary charges $n_{min}(d)$ (see (4)) required to lift a dust particle of diameter $d$, we obtain a useful estimation of the maximum size of the take-off dust grains

$$n_{min}(d) \approx \sigma_\infty(d) \quad \Rightarrow \quad d_{max} \leq \frac{V^{3/10}}{p^{2/5}} 10^{-7} \text{ m}. \qquad (25)$$

Here $V$ is the potential difference inside the double layer (in *volts*), so according to this estimate the upper limit of the size of the take-off dust particles for $V \sim 10$ B is about 100 *nm*.

Finally, using the obtained results, it is easy to calculate the standard deviation for the surface charge density. In the regime of stationary fluctuations, the average number of excess charges $v(d)$ in an area with the square $\pi d^2/4$ is equal to

$$v(d) = 4\sigma_\infty(d)/\pi d^2 = \frac{1}{d}\sqrt{\frac{8 j_{ph}}{\pi e}} \approx \frac{10^7}{d} \text{ m}^{-2}. \qquad (26)$$

This means that the average local density of excess elementary charges $\langle v \rangle$ on the area (**or** dust particle) with a diameter of 100 *nm* is approximately $10^{14}$ $m^{-2}$ and this value will decrease slowly to the value $\langle v \rangle \approx 6 \cdot 10^7$ $m^{-2}$ (1) as the diameter is reduced. The reason is quite obvious: the larger the size of the area over which the averaging is carried out, the larger the number of small areas with the charges different in sign it includes, and the more these charges compensate each other.

### 7. Conclusion

Thus, charge fluctuations play a fundamental important role for work of the electrostatic mechanism of the levitation of dust particles on atmosphereless bodies and, in particular, the levitation of the lunar dust, since just the charge fluctuations lead to an increase of the local density of the surface charge by several orders of magnitude (compare the formulas (1) and (26)). The existence of fluctuations of the surface charge density is the reason for the appearance of a number of effects that are impossible within the framework of the usual "mean field theory":

- the local electric field directly above the surface of the area in which the fluctuation arose is several orders of magnitude stronger than the average field of the double layer $E_{DEL}$;

- as a consequence, the Coulomb forces acting on charged particles that reside directly on the surface or at an altitude up to several tens of nanometers above it are proportionally increasing. Since these powerful forces act at the same characteristic distances as the weak Van der Waals cohesion, the presence of the latter does not affect the levitation of dust above the surface;

- at least one odd or missing electron appears at almost all dust particles superjacent on surface in a few minutes after the start of the lighting;

- the standard deviation of charges of small (with a diameter up to about 100 *nm*) dust particles can be tens of *e* (see (22)), with the result a weak electric field of the double layer $E_{DEL} \sim 1$ *V m$^{-1}$* is capable to lift them above the surface.

- the characteristic lifetime $\tau$ (the time during which the sign of the charge of the fluctuating area changes) is inversely proportional to its size *d* (see (21)) and for the area size $d \sim 100$ *nm* is a few tens of seconds.

- although the rate of all processes is determined by the current density of photoelectrons $j_{ph}$, in the stationary state $\sigma_{\infty}$ depends on $j_{ph}$ only through the potential of the double layer *V,* see (22).

In conclusion, we note that the formulas obtained above determine analytically the dependence of the intensity of the charge fluctuations on various external parameters. Therefore, they can find practical application in the search for technical possibilities for controlling the processes of peeling-off the microparticles adhering to non-conducting surface

**Acknowledgements**

The research was carried out using funds of the Russian scientific foundation (project №17-12-01458).

**References**

Bernstain W., Fredricks R.W., Vogl J.L., Fowler W.A., 1963. The lunar atmosphere and the solar wind, Icarus, 2, 233-248.

Carrier III W.D., Olhoeft G.R., Mendell W., 1991. Physical properties of the Lunar surface, in: Lunar Sourcebook, Eds.: G.H. Heiken, D.T. Vaniman, B.M. French, Cambridge Univ. Press. p. 475, (http://www.lpi.usra.edu/publications/books/lunar_sourcebook/)

Colwell, J.E., Batiste, S., Horanyi, M., Robertson, S., Sture, S., 2007. Lunar surface: Dust dynamics and regolith mechanics, Rev. Geophys. 45, RG2006.

Criswell D.R. and De B.R., 1977. Intense Localized Photoelectric Charging in the Lunar Sunset Terminator Region. 2. Supercharging at the Progression of Sunset, J. Geophys. Res., vol. 82, No. 7, 1005-1007.

De B.R. and Criswell D.R., Intense Localized Photoelectric Charging in the Lunar Sunset Terminator Region. 1. Development of Potentials and Fields, 1977. J. Geophys. Res., vol. 82, No. 7, 999-1004.

Dove A., Horanyi M., Wang X., Piquette M., Poppe A.R., Rebertson S., 2012. Experimental study of a photoelectron sheath, Phys. Plasmas. 19, 043502; doi: 10.1063/1.3700170.

Grobman W.D., Blank J.L., 1969. Electrostatic Potential Distribution of the Sunlit Lunar Surface, J.Geophys.Res.,74, No. 16, 3943-3951.

Feuerbacher, B., Anderegg M., Fitton B., Laude L. D., Willis R. F., Grard R. J. L., 1972. Photoemission from lunar surface fines and the lunar photoelectron sheath. Geochim. Cosmochim. Acta., Suppl. 3, vol. 3, 2655–2663.

Flanagan T. M. and Goree J., Dust release from surfaces exposed to plasma. Physics of plasmas, 2006. **13**, 123504.

Freeman, J. W., Fenner, M. A., Hills, H. K., 1973. The Electric Potential of the Moon in the Solar Wind, J. Geophys. Res., 78, 4560-4567.

Freeman J. W, Ibrahim M., 1975. Lunar electric fields, surface potential and associated plasma sheaths, The Moon, 14, 103-114.

Manka R.H., Plasma and potential at the lunar surface. 1973, in: Grard R.J.L. (Ed.), Photon and Particle Interactions with Surfaces in Space. D. Reidel Publishing Co., Dordrecht, Holland, pp. 347–361,

Manka R.H. Michel F.C., 1973. Lunar ion energy spectra and surface potential, Proceed. of the Fourth Lunar Sci. Conf., (Suppl. 4, Geochimica et Cosmochimica Acta) 3, 2897-2908.

Øksendal B., Stochastic Differential Equations, 2000. Springer-Verlag, Heidelberg, New York.

Opik E.J., Singer S.F., 1960. Escape of gases from the Moon, J. Geophys. Res., 65, 3065-3070.

Popel S.I., Zelenyi L.M., Atamaniuk B., 2016. Plasma in the region of the Lunar terminator, Plasma Physics Reports, 42, No. 5, 543-548.

Poppe A., Horányi M., 2010. Simulations of the photoelectron sheath and dust levitation on the lunar surface, J. Geophys. Res., 115, A08106, doi:10.1029/2010JA015286.

Rosenfeld E.V., 2000. Calculation of the field of a lattice point magnetic dipoles, Physics of the solid state, 42, #9, 1680-1687.

Rosenfeld E.V., Korolev A.V., Zakharov A.V., 2016a. Lunar nanodust: Is it a borderland between powder and gas?, Advances in Space Research 58, 560–563,

Rosenfeld E.V. and Zakharov A.V., 2016b. Role of stochastic processes in particle charging due to photoeffect on the Moon, arXiv:1611.00811 [astro-ph.EP],

Stubbs, T. J., Vondrak R. R., Farrell W. M., 2006. A dynamic fountain model for lunar dust, Adv. Space Res., 37, 59–66, doi:10.1016/j.asr.2005.04.048.

Stubbs T.J., Farrell W.M., Halekas J.S., Burchill J.K., Collier M.R., Zimmerman M.I., Vondrak R.R., Delory G.T., Pfaff R.F., 2014. Dependence of lunar surface charging on solar wind plasma conditions and solar irradiation, Planetary and Space Science, 90, 10–27.

Van Kampen N. G., 2007. Stochastic Processes in Physics and Chemistry, revised and enlarged edition, Elsevier, North-Holland, Amsterdam.

Vaniman B., R. Reedy, G. Heiken, G. Olhoeft, W. Mendell, 1991. The Lunar environment, in: Heiken G.H., Vaniman D.T., French B.M. (Eds.), *Lunar Sourcebook*, *Cambridge Univ. Press,* 27-60 (http://www.lpi.usra.edu/publications/books/lunar_sourcebook/).

Vaverka J., Richterová I., Pavlu J., Šafránková J., Němeček Z., 2016. Lunar surface and dust grain potentials during the Earth`s magnetosphere crossing, The Astrophysical Journal, 825:133ı-10, doi:10.3847/0004-637X/825/2/133ı

Wang, X., Pilewskie J., Hsu H.-W., Horányi M., 2016a. Plasma potential in the sheaths of electron-emitting surfaces in space, Geophys. Res. Lett., 43, 525–531, doi:10.1002/2015GL067175.

Wang, X., Schwan J., Hsu H.-W., Grün E., M. Horányi, 2016b. Dust charging and transport on airless planetary bodies, Geophys. Res. Lett., 43, 6103–6110, doi:10.1002/2016GL069491.

Weiss, G.H.; Rubin, R.J., 1982. Random Walks: Theory and Selected Applications, in: Advances in chemical physics, Prigogine I., Rice S.A. (Eds.), **52**: pp. 363–505, *doi*:*10.1002/9780470142769.ch5*.

Zimmerman M. I., Farrell W. M., Hartzell C. M., Wang X., Horanyi M., Hurley D. M., 2016. Hibbits K., Grain-scale supercharging and breakdown on airless regoliths, American Geophysical Union, doi: 10.1002/2016JE005049.